\documentclass{ws-procs961x669}            
\usepackage[utf8]{inputenc}
\usepackage{enumitem}

\begin{document}
\title{Probing the Primordial Universe using Massive Fields}

\author{Xingang Chen$^{1,2}$, Mohammad Hossein Namjoo$^{1,2}$ and Yi Wang$^{3}$}

\address{
$^1$Institute for Theory and Computation, Harvard-Smithsonian Center for Astrophysics, \\
60 Garden Street, Cambridge, MA 02138, USA \\
$^2$Department of Physics, The University of Texas at Dallas, Richardson, TX 75083, USA\\
$^3$Department of Physics, The Hong Kong University of Science and Technology, \\
  Clear Water Bay, Kowloon, Hong Kong, P.R.China}

\begin{abstract}
Inflation models are numerous. It is extremely difficult, if possible at all, to identify the actual underlying inflation model of our primordial universe. Thus, for the purpose of proving/falsifying inflation and using inflation to probe new physics, model independent approaches are crucial. Massive fields play a uniquely important role in those missions.
This short review is based on a talk by one of the authors (YW) in the 2nd LeCosPA Symposium.
\end{abstract}


\bodymatter

\bigskip

\bigskip\noindent\textit{Introduction.} Inflation is the leading paradigm of the very early universe cosmology. During inflation the universe expands almost exponentially as $a\sim e^{Ht}$, where $H$ is Hubble parameter and is slowly varying. Such a simple scenario explains horizon and flatness puzzles in the hot big bang cosmology and predicts primordial density fluctuations which turn out to be the seeds of the cosmic microwave background (CMB) anisotropies and the large scale structure (LSS).

What would be a logical route to systematically investigate the inflationary physics? A naïve approach would be copying what collider particle physicists have done.

In particle physics, building on the general principles of quantum field theory, the particle physics standard model is established in the 1960s. In the 50 years that follows, the standard model is tested to agree with experiments extremely well. The standard model now provides a reference model for the search of new physics, which is referred to as beyond standard model physics.

If the similar thing would about to happen for inflation, one would first build inflation models, and afterwards test those models and pick the right one from experiments. Once the standard model of inflation is established, one can then use the model as a standard reference to probe new physics during inflation.

Unfortunately, such an approach for inflation fails badly, simply because there are too many inflation models, way too many compared to the number of observables that one typically expects to observe.

For example, in Encyclopædia Inflationaris \cite{Martin:2013tda}, 74 models (potentials) of single field slow roll inflation are summarized. Even before addressing the completeness of this list, one can easily come up with inflation models with a summation of some those potential terms resulting in a combinatorial number of new models. On top of that, one can add scale dependent sharp features \cite{Starobinsky:1992ts, Adams:2001vc, Chen:2006xjb, Chen:2008wn}, non-Bunch-Davies initial conditions \cite{Easther:2001fi, Chen:2006nt, Jiang:2015hfa, Jiang:2016nok}, generalized kinetic terms such as Galileons \cite{Kobayashi:2010cm}, and so on and so forth.

This is more or less the story for single scalar field inflation. One can in addition consider quasi-single field \cite{Chen:2009we, Chen:2009zp, Baumann:2011nk}, multi-field, and generalize to higher spin fields \cite{Golovnev:2008cf}. As a result, the possibilities of inflation models is practically infinite. (Beyond those practically infinite possibilities of inflation, there are also all kinds of alternative to inflation scenarios, which we shall address later.) Although lots of experiments are on going or being planned, we are still very unlikely to pin down a standard model for inflation. So the attempt to copy the theoretical development of particle physics very likely leads to an unfortunate end here.

Why such a  ``logical'' route fails? This is related to the simple $a\sim e^{Ht}$ nature of inflation. Ugly, bad and good consequences arise from here. As the ugly consequence, this is why there are so many inflationary models. Any potential satisfying slow roll conditions can inflate. Although WMAP and Planck marks the era of ruling out simple models such as $\lambda \phi^4$, the possibility is still widely open for many models mentioned above. As the bad consequence, those models predict very similar observational signatures, which only differ by $1\%$ (the order of slow roll parameters) or smaller. This adds to the extreme difficulty in distinguishing inflationary models observationally.

Fortunately, there also exists a good consequence: As long as one postulates $a\sim e^{Ht}$, model independent results can be obtained and one can use such model independent results to test inflation in a model independent way, and probe new high energy physics model independently based on the scenario of inflation, instead of a particular inflation model.

A nice example -- and the only example for a long time in the literature -- is gravitational waves \cite{Starobinsky:1979ty}. Gravitational waves is the universal prediction of inflation. Once observed, it is a non-trivial test of inflation, and indicates new physics -- the quantum fluctuation of the metric degree of freedom.

Recently, another example of such is realized -- the massive fields during inflation \cite{Chen:2009we, Chen:2009zp, Baumann:2011nk, Chen:2012ge, Arkani-Hamed:2015bza, Chen:2011zf, Chen:2014cwa, Chen:2015lza}. The massive fields predict unique signature in observations. Once observed, the expansion history of the inflationary universe can be reconstructed and thus it is a direct probe of inflation. Moreover, the detection of new massive particles during inflation is an invaluable hint for the development of high energy physics.

\bigskip\noindent\textit{Massive fields during inflation.}
The inflaton field is not the only field existing during inflation. There exists plenty of massive fields during inflation. Those massive fields can come from:
\begin{itemize}[leftmargin=0.5cm]
  \item IR uplifting. There are many fields in the particle physics standard model. The flat space mass spectrum of those fields are negligibly small in the context of a typical inflation model. However, on an inflationary background, those fields get masses due to the Gibbons-Hawking temperature $M\sim g T\sim g H/(2\pi)$, where $g$ is a coupling constant (or its square root) depending on the details of the particle.
  \item UV completion. A quantum field theory with UV divergence needs UV completion. Inflation especially requires UV completion because even a dimension 6 operator $V(\phi)\phi^2/M_p^2$ contributes part of the inflaton mass $\delta m \sim H$ and thus spoils the slow roll potential. In the viewpoint of effective field theory, new massive modes arise at the UV completion scale. For example, in string theory, there are various massive modes coming from moduli spaces, compactifications, stringy excitations, and so on. Those massive modes are typically below or at the string scale.
  \item Supersymmetry breaking. It is often believed that the inflationary energy scale is high and there is supersymmetry at such high energy scales. However, supersymmetry must be broken during inflation because of the quasi-de Sitter geometry. Thus the light fields whose mass is protected by supersymmetry obtain a correction of order Hubble \cite{Copeland:1994vg, Baumann:2011nk}.
\end{itemize}

In the long wavelength limit, a massive field oscillates as $(-\tau)^{\pm i \mu}$, where $\tau$ is the conformal time, and $\mu \equiv \sqrt{M^2/H^2-\alpha^2}$, where $\alpha$ depends on spin. For spin 0, 1/2 and 1, $\alpha$ takes value 3/2, 0 and 1/2 respectively. When $M/H<\alpha$ (which applies only for scalars and vectors), the massive field behaves as an over-damped oscillator and shall not oscillate in the IR. However, when $M/H>\alpha$, the Hubble friction cannot stop the field from oscillating. And fermions oscillate in the IR regardless of their mass. We focus on the $M/H>\alpha$ case in the following, and the results of the $M/H<\alpha$ case can be obtained by an analytical continuation.

The contributions from the massive fields, imprinting onto the primordial density fluctuations, can be classified into two types, namely local and non-local ones \cite{Arkani-Hamed:2015bza}. The non-local contribution originates from thermal particle production and cannot be be integrated out or mimicked by a single inflaton effective theory. The comparison between the local and non-local contribution is shown in Table \ref{tab:locality}.

\begin{figure}[htbp]
  \begin{center}\begin{tabular}{ | c | c | c | c | c |}
    \hline
    name      & origin  & analyticity  & integrate out & size
    \\ \hline\hline
    local     & vacuum  & analytic     & can           & $1/\mu^2$
    \\ \hline
    non-local & thermal & non-analytic & cannot        & $e^{-\pi \mu}$
    \\ \hline
  \end{tabular}\end{center}
  \caption{\label{tab:locality} Comparison of different imprints on density fluctuation from a massive field.}
\end{figure}

\noindent\textit{Detecting new massive particles in the cosmological collider \cite{Chen:2009we, Chen:2009zp, Baumann:2011nk, Arkani-Hamed:2015bza}.}
The key mission for collider particle physics is searching for new particles. The major technique of new particle detection is through its resonance peaks. Things are similar here for the primordial universe. An under-damped massive field oscillates as $\sigma\sim \exp(i M t)$. Thus the three point function between this field and two curvature scalar fields contains
\begin{align}
  \label{eq:3pt}
  \left \langle \zeta_ \mathbf{k}^2 \sigma \right \rangle
  \supset
  \int f(\tau) e^{i (Mt -2 k \tau)} d\tau~
  \propto
  e^{i (Mt_* - 2k\tau_*)} ~,
\end{align}
where $f(\tau)$ is a slowly varying function of time. As $\tau_*$ (or $t_*$) satisfying $k/a(\tau_*) = M/2$, a resonance is triggered and the density fluctuation is amplified \cite{Chen:2008wn}; and we have used the saddle point approximation in the second step. The effect of the $\sigma$ field converts to density fluctuations either through gravitational or direct coupling. At the time of resonance, the phase of the massive field quantum oscillation is recorded in the above integral.
This oscillatory behavior in $k$ is the characteristic signal from massive fields. In single field effective field theory, we expect the correlation functions to be analytic in $k$ thus such terms cannot appear.
Note that the cases of $M/H<\alpha$ and $M/H>\alpha$ are related by an analytic continuation.

\bigskip\noindent\textit{Towards proving/falsifying the inflation scenario \cite{Chen:2011zf, Chen:2014cwa, Chen:2015lza}.}
It is interesting to view the $\left \langle \zeta_ \mathbf{k}^2 \sigma \right \rangle$ correlation in a broader sense. While inflation is the leading paradigm of the primordial universe, there are possible alternatives. The massive fields can be considered as a clock, whose ticks and tocks are uniform in physical time and labelling the different modes in density perturbations with its phases. Depending on whether this oscillation of massive fields is classical or quantum-mechanical, we have the classical or quantum standard clocks, respectively.
Solving the resonance condition $a(t_*) = 2k/M$ for $t_*$ or $\tau_*$, one can see that the phase of the correlation function \eqref{eq:3pt} as a function of k is proportional to the inverse function of $a(t_*)$ or $a(\tau_*)$, so the evolution of the scale factor of the primordial universe is directly recorded in the astrophysical observables.

\bigskip\noindent\textit{Summary and discussions.} We would like to conclude this proceeding paper by comparing the physical implication of inflationary massive fields with inflationary gravitational waves, because unlike the other observables both of them can be used to distinguish the inflation from alternative scenarios model-independently. We refer them as ``standard clock signals" versus ``gravitational waves":

\begin{itemize}[leftmargin=0.5cm]
  \item Both exist in Nature model-independently. For standard clock signals, we refer to the quantum standard clock signal \cite{Chen:2015lza} in this statement.

  \item Both predict characteristic observational features.
  \begin{itemize}[leftmargin=0.5cm]
    \item The gravitational waves are imprinted on the CMB as the B-mode polarization, which is hard to mimic by other primordial physics (though not impossible, for example, primordial non-decaying vector fluctuations).
    \item The standard clock signals are imprinted in all kinds of density fluctuations as special types of oscillatory features.
        Classical standard clocks predict unique scale-dependent oscillatory signals in power spectrum and correlatedly in non-Gaussianites. Quantum standard clocks predict shape-dependent oscillatory signals in non-Gaussianities, which are protected by the analyticity of quantum field theory.
  \end{itemize}

  \item Both test the inflation scenario in a model independent way, with some caveats.
  \begin{itemize}[leftmargin=0.5cm]
    \item The conventional wisdom is that, the amplitude $H^2/M_p^2$ (and its tilt) of the gravitational wave directly probes the Hubble scale (and its time dependence), and by observing a scale-invariant gravitational waves one proves inflation. However, there are a few caveats. First, conservation of the amplitude of the gravitational wave outside the horizon is assumed. However, this is not true in contracting universes, where the growing mode dominates over the constant mode. As a result, matter contraction, \cite{Wands:1998yp, Finelli:2001sr} cannot be distinguished from inflation by the amplitude and tilt of the gravitational waves. Moreover, non-trivial time evolution (even parametric resonance) of the super-horizon gravitational waves can also arise in massive gravity \cite{Lin:2015nda}. Second, vacuum initial condition is assumed, and by relaxing it the string gas cosmology \cite{Brandenberger:1988aj} may produce the same amplitude (and see \cite{Wang:2014kqa, blueprep} for a discussion of tilt). Moreover, non-Bunch Davies initial conditions or sources from the matter sector (for example, particle production \cite{Mukohyama:2014gba} or enhanced non-linear perturbations \cite{Biagetti:2013kwa}) can also change the amplitude and tilt of the gravitational waves. So among the variety of alternative-to-inflation scenarios in the literature, the gravitational wave mainly distinguishes inflation from the ekpyrosis scenario \cite{Khoury:2001wf}. Even between inflation and ekpyrosis, it is worth noting that the gravitational wave is a one-sided test -- seeing gravitational waves supports inflation and rules out ekpyrosis but not seeing gravitational waves shall not support ekpyrosis or rule out inflation.
    \item The standard clock signal is model independent. The oscillation frequency does not depend on initial condition or super-horizon evolution of the field (though the amplitude depends on those details and is model dependent). It is shown that the clock signal distinguishes the inflation, ekpyrosis, string gas and matter bounce scenarios by predicting a unique signal pattern for each scenario.
        The frequency of the massive field oscillation can be affected by time dependent parameters in the Lagrangian. A fine tuned $m(t)$ may let the clock signal record a deformed evolution history of the primordial universe. These are interesting topics for future investigations.
  \end{itemize}

  \item Both indicate significant new physics if detected.
  \begin{itemize}[leftmargin=0.5cm]
    \item Gravitational waves indicate the quantized fluctuations of gravity. Although in theory its existence is expected, an actual discovery would pin down the energy scale of inflation, which is new physics.
    \item Standard clock signals is a probe of massive fields. Similarly, although in theory their existence is expected, an actual discovery would tell us the detailed information about the particle spectrum and coupling, along with, perhaps even more importantly, the recorded evolutionary history of the primordial universe.
  \end{itemize}

  \item Unfortunately, both are hard to detect.
  \begin{itemize}[leftmargin=0.5cm]
    \item The amplitude of gravitational waves is of order $H^2/M_p^2$. The Planck suppression make it hard to detect, especially for low scale inflation models \cite{Graham:2015cka, Jiang:2015qor}. Even worse, the gravitational waves behaves as radiation and decays after returning to the horizon. Thus we have to rely on observations on large scales, and the number of modes is limited. Also, there can be contamination from astrophysical sources such as polarized emissions from dust.
    \item The quantum standard clock signals can be suppressed by two factors, the mass and the coupling. When the mass of the field is large, the suppression factor is $\exp(-\pi M/H)$.\cite{Arkani-Hamed:2015bza} Thus probing very massive fields is exponentially hard. Although it's natural to expect fields with $M\sim H$, we do not know it for sure. Coupling-wise, direct couplings generically predict much larger signals than gravitational coupling, giving rise to $f_{NL}\sim \epsilon$ or much larger.\cite{Chen:2009zp,Baumann:2011nk} These signals may be observable in future experiments. The classical standard clock signals are easier to observe but its existence is model-dependent. Observation-wise, the squeezed limit non-Gaussianity gives the sharpest clock signals, but can only make use of a small subset of available modes. On the other hand, oscillatory signals may be more difficult to analyze but less susceptible to contaminations from astrophysical sources.
  \end{itemize}
\end{itemize}

As well-known, the primordial gravitational waves is a very important probe of the scenario type of the primordial universe. On the other hand, recent researches have uncovered another set of phenomena induced by massive fields, which could be used to achieve the same and complimentary science goals. The study is in its early stage and there are many works need to be done.

\bigskip\noindent\textit{Acknowledgements} YW thanks organizers of the 2nd LeCosPA Symposium, especially Pisin Chen and Yifu Cai for invitation. XC and MHN are supported in part by the NSF grant PHY-1417421. YW is supported by the CRF Grants of the Government of the Hong Kong SAR under HKUST4/CRF/13G.

\bibliographystyle{ws-procs961x669}
\bibliography{QPSC}

\begin{thebibliography}{10}

\bibitem{Martin:2013tda}
J.~Martin, C.~Ringeval and V.~Vennin, {Encyclopædia Inflationaris}, {\em Phys.
  Dark Univ.} {\bf 5-6}, 75  (2014).

\bibitem{Starobinsky:1992ts}
A.~A. Starobinsky, {Spectrum of adiabatic perturbations in the universe when
  there are singularities in the inflation potential}, {\em JETP Lett.} {\bf
  55}, 489  (1992), [Pisma Zh. Eksp. Teor. Fiz.55,477(1992)].

\bibitem{Adams:2001vc}
J.~A. Adams, B.~Cresswell and R.~Easther, {Inflationary perturbations from a
  potential with a step}, {\em Phys. Rev.} {\bf D64}, p. 123514  (2001).

\bibitem{Chen:2006xjb}
X.~Chen, R.~Easther and E.~A. Lim, {Large Non-Gaussianities in Single Field
  Inflation}, {\em JCAP} {\bf 0706}, p. 023  (2007).

\bibitem{Chen:2008wn}
X.~Chen, R.~Easther and E.~A. Lim, {Generation and Characterization of Large
  Non-Gaussianities in Single Field Inflation}, {\em JCAP} {\bf 0804}, p. 010
  (2008).

\bibitem{Easther:2001fi}
R.~Easther, B.~R. Greene, W.~H. Kinney and G.~Shiu, {Inflation as a probe of
  short distance physics}, {\em Phys. Rev.} {\bf D64}, p. 103502  (2001).

\bibitem{Chen:2006nt}
X.~Chen, M.-x. Huang, S.~Kachru and G.~Shiu, {Observational signatures and
  non-Gaussianities of general single field inflation}, {\em JCAP} {\bf 0701},
  p. 002  (2007).

\bibitem{Jiang:2015hfa}
H.~Jiang and Y.~Wang, {Towards the physical vacuum of cosmic inflation}
  (2015).

\bibitem{Jiang:2016nok}
H.~Jiang, Y.~Wang and S.~Zhou, {On the initial condition of inflationary
  fluctuations}  (2016).

\bibitem{Kobayashi:2010cm}
T.~Kobayashi, M.~Yamaguchi and J.~Yokoyama, {G-inflation: Inflation driven by
  the Galileon field}, {\em Phys. Rev. Lett.} {\bf 105}, p. 231302  (2010).

\bibitem{Chen:2009we}
X.~Chen and Y.~Wang, {Large non-Gaussianities with Intermediate Shapes from
  Quasi-Single Field Inflation}, {\em Phys. Rev.} {\bf D81}, p. 063511  (2010).

\bibitem{Chen:2009zp}
X.~Chen and Y.~Wang, {Quasi-Single Field Inflation and Non-Gaussianities}, {\em
  JCAP} {\bf 1004}, p. 027  (2010).

\bibitem{Baumann:2011nk}
D.~Baumann and D.~Green, {Signatures of Supersymmetry from the Early Universe},
  {\em Phys. Rev.} {\bf D85}, p. 103520  (2012).

\bibitem{Golovnev:2008cf}
A.~Golovnev, V.~Mukhanov and V.~Vanchurin, {Vector Inflation}, {\em JCAP} {\bf
  0806}, p. 009  (2008).

\bibitem{Starobinsky:1979ty}
A.~A. Starobinsky, {Spectrum of relict gravitational radiation and the early
  state of the universe}, {\em JETP Lett.} {\bf 30}, 682  (1979), [Pisma Zh.
  Eksp. Teor. Fiz.30,719(1979)].

\bibitem{Chen:2012ge}
X.~Chen and Y.~Wang, {Quasi-Single Field Inflation with Large Mass}, {\em JCAP}
  {\bf 1209}, p. 021  (2012).

\bibitem{Arkani-Hamed:2015bza}
N.~Arkani-Hamed and J.~Maldacena, {Cosmological Collider Physics}  (2015).

\bibitem{Chen:2011zf}
X.~Chen, {Primordial Features as Evidence for Inflation}, {\em JCAP} {\bf
  1201}, p. 038  (2012).

\bibitem{Chen:2014cwa}
X.~Chen, M.~H. Namjoo and Y.~Wang, {Models of the Primordial Standard Clock},
  {\em JCAP} {\bf 1502}, p. 027  (2015).

\bibitem{Chen:2015lza}
X.~Chen, M.~H. Namjoo and Y.~Wang, {Quantum Primordial Standard Clocks}
  (2015).

\bibitem{Copeland:1994vg}
E.~J. Copeland, A.~R. Liddle, D.~H. Lyth, E.~D. Stewart and D.~Wands, {False
  vacuum inflation with Einstein gravity}, {\em Phys. Rev.} {\bf D49}, 6410
  (1994).

\bibitem{Wands:1998yp}
D.~Wands, {Duality invariance of cosmological perturbation spectra}, {\em Phys.
  Rev.} {\bf D60}, p. 023507  (1999).

\bibitem{Finelli:2001sr}
F.~Finelli and R.~Brandenberger, {On the generation of a scale invariant
  spectrum of adiabatic fluctuations in cosmological models with a contracting
  phase}, {\em Phys. Rev.} {\bf D65}, p. 103522  (2002).

\bibitem{Lin:2015nda}
C.~Lin and M.~Sasaki, {Resonant Primordial Gravitational Waves Amplification},
  {\em Phys. Lett.} {\bf B752}, 84  (2016).

\bibitem{Brandenberger:1988aj}
R.~H. Brandenberger and C.~Vafa, {Superstrings in the Early Universe}, {\em
  Nucl. Phys.} {\bf B316}, p. 391  (1989).

\bibitem{Wang:2014kqa}
Y.~Wang and W.~Xue, {Inflation and Alternatives with Blue Tensor Spectra}, {\em
  JCAP} {\bf 1410}, p. 075  (2014).

\bibitem{blueprep}
M.~He, J.~Liu, S.~Lu, S.~Zhou, Y.~Cai, Y.~Wang and R.~Brandenberger, {To
  appear}.

\bibitem{Mukohyama:2014gba}
S.~Mukohyama, R.~Namba, M.~Peloso and G.~Shiu, {Blue Tensor Spectrum from
  Particle Production during Inflation}, {\em JCAP} {\bf 1408}, p. 036  (2014).

\bibitem{Biagetti:2013kwa}
M.~Biagetti, M.~Fasiello and A.~Riotto, {Enhancing Inflationary Tensor Modes
  through Spectator Fields}, {\em Phys. Rev.} {\bf D88}, p. 103518  (2013).

\bibitem{Khoury:2001wf}
J.~Khoury, B.~A. Ovrut, P.~J. Steinhardt and N.~Turok, {The Ekpyrotic universe:
  Colliding branes and the origin of the hot big bang}, {\em Phys. Rev.} {\bf
  D64}, p. 123522  (2001).

\bibitem{Graham:2015cka}
P.~W. Graham, D.~E. Kaplan and S.~Rajendran, {Cosmological Relaxation of the
  Electroweak Scale}, {\em Phys. Rev. Lett.} {\bf 115}, p. 221801  (2015).

\bibitem{Jiang:2015qor}
H.~Jiang, T.~Liu, S.~Sun and Y.~Wang, {Echoes of Inflationary Particle Phase
  Transitions in the CMB}  (2015).

\end{thebibliography}

\end{document}